**Special spin behavior of rare earth ions at the A site of polycrystalline ErFe$_{1-x}$Cr$_x$O$_3$ (x = 0.1, 0.9)**

Jiyu Shen[1], Jiajun Mo[1], Zeyi Lu[1], Zhongjin Wu[1], Chenying Gong[1], Kaiyang Gao[1], Pinglu Zheng[2], Min Liu*[,1], Yanfang Xia*[,1]

[1]College of Nuclear Science and Technology, University of South China, Hengyang 421200, Hunan, P.R China.

[2]Information Engineering Institute, Wenzhou Business College, Wenzhou 325000, Zhejiang, P.R China.

Email: liuhart@126.com, xiayfusc@126.com

**Abstract**

Thermally induced spin control is one of the main directions for future spin devices. In this study, we synthesized single-phase polycrystalline ErFe$_{1-x}$Cr$_x$O$_3$ and combined the magnetization curves and Mössbauer spectra to determine the macroscopic magnetism at room temperature. The magnetization of the system at various temperatures is well simulated by molecular field theory. And it is found that under the DM interaction, not only the B-site ions undergo a reorientation process, but the spins of the A-site ions also change at the same time. The effective spin is defined as the projection of Er$^{3+}$ on the Fe$^{3+}$/Cr$^{3+}$ spin plane, and the whole reorientation process is obtained by fitting. This study will complement the actual process of ErFe$_{1-x}$Cr$_x$O$_3$ spin reorientation and will lay a theoretical foundation for the fabrication of future spin-controlled devices.

1. Introduction

Rare earth perovskites have maintained a high research interest for a long time, only because of their special and complex magnetic behaviour [1-3] and strong application potential [4-6]. As pointed out in the classic review by White [7], the studied compounds exhibit enormous and sometimes puzzling magnetic properties.

Unlike general perovskites, when magnetic rare earth ions (such as Pr, Nd, Sm, Eu, Gd, Tb, Dy, Ho, Er, Tm, Yb, and Lu) are introduced, the superexchange of the system will become very complicated. However, current researchers are not satisfied with the performance study of a single parent, and often like to change the properties of some aspects of the system by doping. At this time, there will be six exchange effects in the system. Not only that, special properties such as Dzyaloshinskii-Moriya (DM) interaction, anisotropy, spin reorientation, etc. will also have a huge impact on the overall magnetic properties of the system [8-10].

$ErFe_{1-x}Cr_xO_3$ is a typical rare earth perovskite system with Curie temperatures around 130 K ($x = 1$) to 625 K ($x = 0$). So far, many researchers have carried out many excellent studies on its complex magnetic properties. Ramu et al. [11] studied the special magnetic moment reversal phenomenon of $ErFeO_3$ at low temperature; Katba et al. [12] changed the magnetic properties of $EeFeO_3$ by doping non-magnetic La in the A site; Massa et al. [13] explored the A site by terahertz spectroscopy Effects on magnetoelectric dynamics. Nevertheless, there are still many unknown mechanisms for $ErFe_{1-x}Cr_xO_3$ rare earth perovskites worthy of discovery.

Spin reorientation is a special magnetic field in magnetic rare-earth perovskites.

Zubov et al. [14] studied the effect of DM interaction and Re-Fe exchange anisotropy on reorientation temperature and compensation temperature by means of mean field theory. Shen et al. [15] investigated the continuous and discontinuous processes of low-temperature spin reorientation of $Fe^{3+}$ in single crystal $ErFeO_3$ by AC susceptibility measurements at various frequencies and static magnetic fields. Ma et al. [16] achieved low-field controlled spin switching of single-crystal $ErFeO_3$. It is not difficult to conclude, however, that researchers generally believe that only B-site ions can redirect. However, by calculating and referring to previous articles [17], we found that a certain redirection process also occurs in A-site $Er^{3+}$.

Therefore, in this report, we will determine the macroscopic magnetic properties at a fixed temperature by means of Mössbauer spectra and magnetization curves, and use the molecular field model to simulate the magnetization curve of $ErFe_{1-x}Cr_xO_3$ in the long-range temperature range. The effective spin is defined as the projection of $Er^{3+}$ on the $Fe^{3+}/Cr^{3+}$ spin plane, and then the optimal curve is obtained by fitting it. This work will complement the entire process of spin reorientation in erbium-based rare earth perovskites, laying the foundation for the fabrication of future ultrafast precise spin devices.

## 2. Experimental

2.1. Sample synthesis

$ErFe_{1-x}Cr_xO_3$ ($x$ = 0.1, 0.9) polycrystalline rare earth perovskites are prepared by a simple sol-gel combustion method. All the precursors required for the synthesis were obtained from MackLin, namely Erbium nitrate, Ferric nitrate, Chromium nitrate,

Ethylene glycol, Citric acid. All nitrates were dissolved in deionized water according to a certain ion ratio ($Er^{3+}$: $Fe^{3+}$: $Cr^{3+}$ = 1: 1-$x$: $x$, $x$=0.1, 0.9). Heating with continuous stirring at 80 °C, and adding an appropriate amount of ethylene glycol and citric acid in the process, the molar ratio of the two is 1:1, until a gel is formed. The gel was then combusted at 150°C to form a honeycomb black/brown solid. Finally, it was pre-calcined at 600°C for 12 hours and then calcined at 1200°C for a further 24 hours.

2.2. Sample Characterization

2.2.1. X-ray diffraction measurements (XRD)

X-ray diffraction (XRD) experiments were carried out in Siemens D500 Cu Kα (λ = 1.5418 Å) diffractometer with the range of 20° to 80° (rate of 0.02°/s). The obtained data were processed with Fullprof software.

2.2.2. Mössbauer spectrum Test

The transmission $^{57}$Fe Mössbauer spectra of $ErFe_{0.9}Cr_{0.1}O_3$ and $ErFe_{0.1}Cr_{0.9}O_3$ were collected at room temperature(RT) on SEE Co W304 Mössbauer spectrometer with a $^{57}$Co/Rh source in transmission geometry equipped in a cryostat (Advanced Research Systems, Inc.,4 K). The data results were fitted with MössWinn 4.0 software.

2.2.3. Magnetic Test

The thermomagnetic curves (M-T) of the samples $ErFe_{0.9}Cr_{0.1}O_3$ and $ErFe_{0.1}Cr_{0.9}O_3$ were obtained under an external magnetic field of 100 Oe. Field-cooled (FC) and zero-field-cooled (ZFC) measurements were done at a magnetic field strength of 100 Oe and temperatures ranging from 10 to 400 K. The magnetization curves (M-H) are obtained at temperatures of 300 K.

## 3. Results and discussions

3.1. XRD analysis

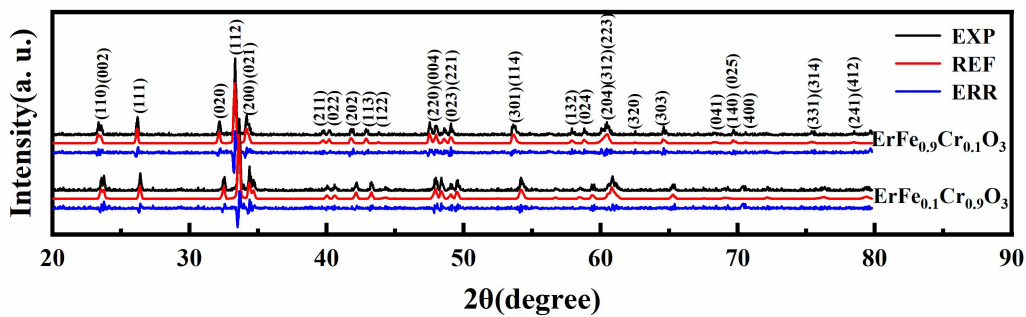

**Fig. 1.** X-ray diffraction spectrum of ErFe$_{0.9}$Cr$_{0.1}$O$_3$ and ErFe$_{0.1}$Cr$_{0.9}$O$_3$ (refined with Fullprof software).

**Table. 1.** Lattice parameters, unit cell volume and other parameters.

| Sample | a (Å) | b (Å) | c (Å) | Cell volume (Å³) | Average grain size (Å) | Bulk density (g/cm³) |
|---|---|---|---|---|---|---|
| ErFe$_{0.9}$Cr$_{0.1}$O$_3$ | 5.263 | 5.582 | 7.591 | 223.01 | 652 | 8.0741 |
| ErFe$_{0.1}$Cr$_{0.9}$O$_3$ | 5.223 | 5.516 | 7.519 | 216.62 | 615 | 8.1940 |

The XRD pattern of ErFe$_{0.9}$Cr$_{0.1}$O$_3$ and ErFe$_{0.1}$Cr$_{0.9}$O$_3$ are shown in Figure 1. Datas and errors refined by Fullprof are shown in red and blue, respectively. All major Bragg diffraction peaks are marked in the figure. From the figure, it can be observed that both samples belong to the orthorhombic system, and the space group is *Pbnm*. And the existence of impurity phase was not obviously observed.

The lattice parameters after ErFe$_{0.9}$Cr$_{0.1}$O$_3$ and ErFe$_{0.1}$Cr$_{0.9}$O$_3$ refinement are shown in Table 1. By comparison, we found that when the iron content increased, both the lattice parameters, the unit cell volume and the average grain size showed an increasing phenomenon. This is attributed to the substitution of larger radius ions, ie $Fe^{3+}$(0.645 Å) → $Cr^{3+}$(0.615 Å) [18].

It was observed that the XRD lines of the multiferroic samples shifted significantly to low angles. This phenomenon, which is very common in polycrystalline semiconductor compounds and also appeared in our previous work [19], is due to increased stress gradients. The crystal stress is expressed by the Williamson-Hall (W-H) method, i.e., formula (1) [20,21]:

$$\beta \cos \theta = k\lambda/D + 4\varepsilon \cos \theta \quad \#(1)$$

Where $k = 0.9$, $\lambda$ is the X-ray wavelength (1.5418 nm), $\beta$ is the full width at half maximum of the peak, $D$ is the average crystallite size, $\theta$ is the Bragg angle, and $\varepsilon$ is the macrostrain.

The tolerance factor $t$ is often used to represent the stability and distortion of the perovskite lattice, and is expressed by formula (2) [22]:

$$t = \frac{R_A + R_O}{\sqrt{2}(R_B + R_O)} \quad \#(2)$$

Where $R_A$, $R_B$, and $R_O$ represent the ionic radii of A site, B site and oxygen, respectively. For $R_B$:

$$R_B = (1-x)R_{Fe^{3+}} + xR_{Cr^{3+}} \quad \#(3)$$

It is not difficult to see that with the increase of Fe, the tolerance factor t shows signs of decreasing, and the possibility of reflecting lattice distortion increases. This is also the source of the orthogonal structure of the $ErFe_{1-x}Cr_xO_3$ system.

3.2. Magnetization curve (M-H) and Mössbauer analysis

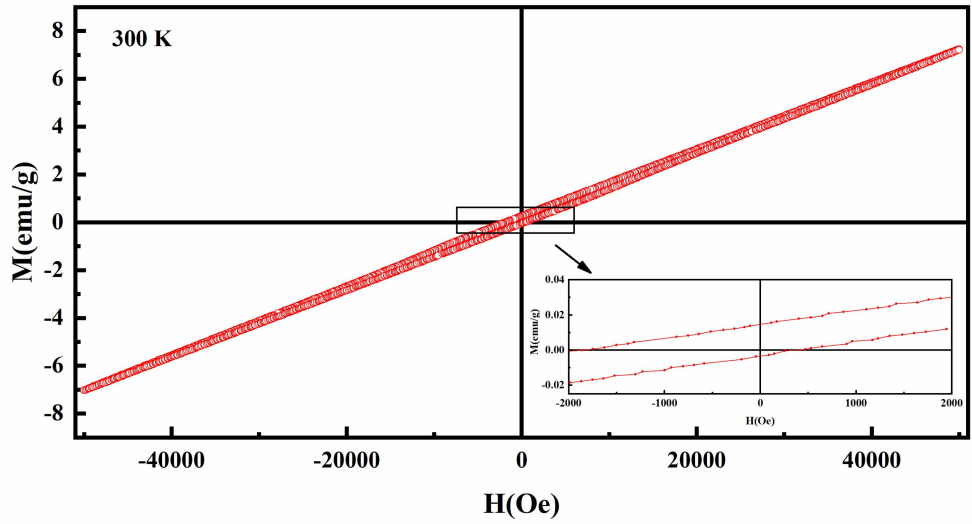

**Fig. 2.** Magnetization curves of ErFe$_{0.9}$Cr$_{0.1}$O$_3$ at a temperature of 300 K, the inset is an enlarged view of -0.2 T-0.2 T.

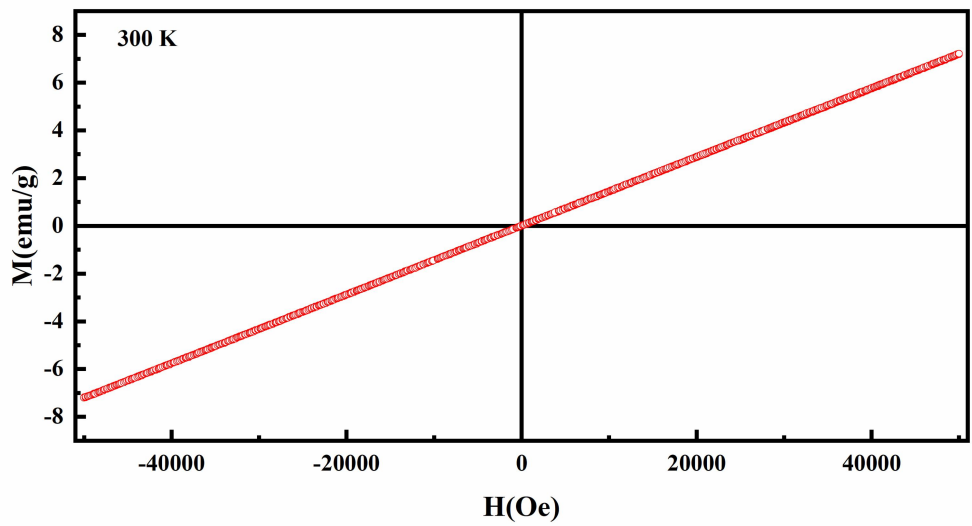

**Fig. 3.** Magnetization curves of ErFe$_{0.1}$Cr$_{0.9}$O$_3$ at a temperature of 300 K.

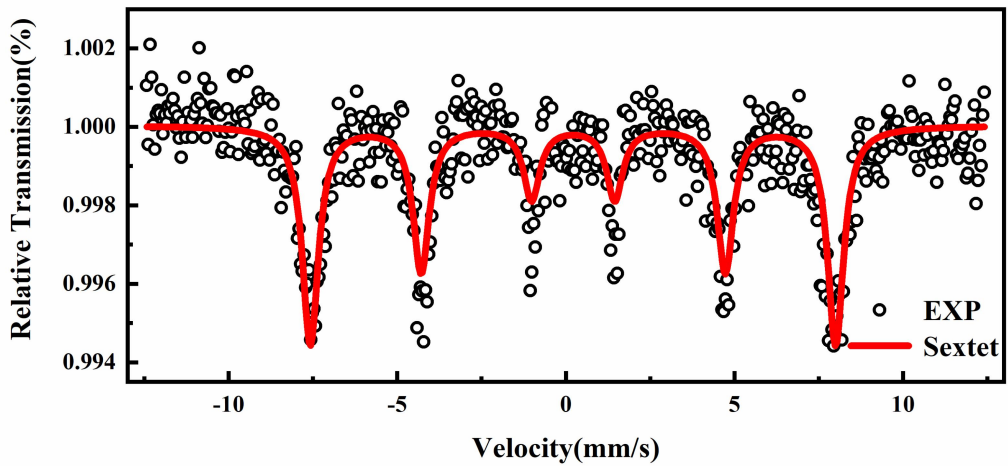

**Fig. 4.** Mössbauer spectrum of ErFe$_{0.9}$Cr$_{0.1}$O$_3$ at 300 K.

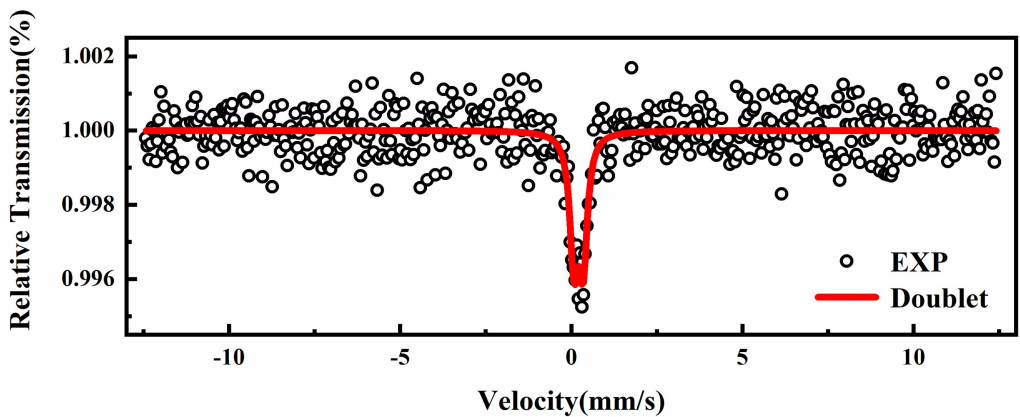

**Fig. 5.** Mössbauer spectrum of ErFe$_{0.1}$Cr$_{0.9}$O$_3$ at 300 K.

Figures 2 and 3 are the magnetization curves of ErFe$_{0.9}$Cr$_{0.1}$O$_3$ and ErFe$_{0.1}$Cr$_{0.9}$O$_3$ at 300 K, respectively. A distinct hysteresis loop can be seen in ErFe$_{0.9}$Cr$_{0.1}$O$_3$. Not only that, the magnetization also increases linearly at a certain rate under increasing external field, and has no tendency to saturate under high field. This is an obvious weak ferromagnetic phenomenon. That is, the spins of Fe$^{3+}$/Cr$^{3+}$ ions do not exhibit antiparallel alignment, but have a certain angle, resulting in a small contribution to the net magnetic moment. In the data of ErFe$_{0.1}$Cr$_{0.9}$O$_3$, no obvious hysteresis loop was observed. Not only that, but we also found a strong exchange bias effect in ErFe$_{0.9}$Cr$_{0.1}$O$_3$, that is, the magnetization curve deviates from the center point symmetry. In past studies, this phenomenon is usually attributed to the competition between ferromagnetism and antiferromagnetism [23].

Figures 4 and 5 show the Mössbauer spectra of ErFe$_{0.9}$Cr$_{0.1}$O$_3$ and ErFe$_{0.1}$Cr$_{0.9}$O$_3$ at 300 K, and Table 2 shows the hyperfine parameters obtained by fitting. Both exhibited the presence of a single Fe component (ie, only one spectral line was present). The difference is that a single sextet is resolved in ErFe$_{0.9}$Cr$_{0.1}$O$_3$, while a single doublet is resolved in ErFe$_{0.1}$Cr$_{0.9}$O$_3$. correspond to the weak ferromagnetic and paramagnetic properties, respectively, which is the same conclusion as the magnetization curve. Comparing the values of QS and IS, we can confirm the existence of trivalent high-spin Fe.

**Table. 2.** Hyperfine parameters of ErFe$_{0.9}$Cr$_{0.1}$O$_3$ and ErFe$_{0.1}$Cr$_{0.9}$O$_3$.

| Sample | QS(mm/s) | IS(mm/s) | H(T) | Γ (mm/s) |
|---|---|---|---|---|
| ErFe$_{0.9}$Cr$_{0.1}$O$_3$ | -0.007 | 0.215 | 48.422 | 0.582 |
| ErFe$_{0.1}$Cr$_{0.9}$O$_3$ | 0.268 | 0.215 | / | 0.289 |

QS: quadruple splitting, IS: isomer shift, Γ: Lorentzian linewidth, H: hyperfine magnetic field.

3.3. Thermomagnetic Curve Analysis and Simulation

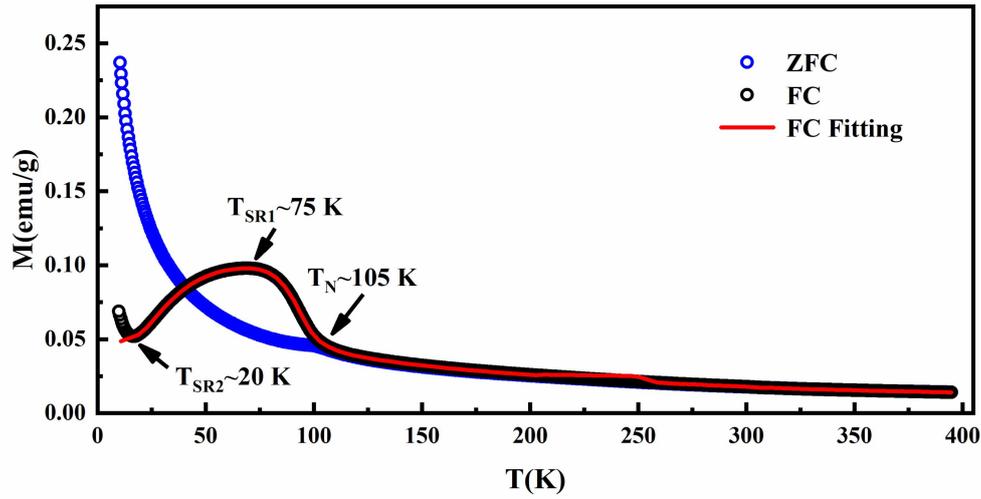

**Fig. 6.** Thermomagnetic curves of ErFe$_{0.9}$Cr$_{0.1}$O$_3$ at an external field of 100 Oe, and fitted with a molecular field model.

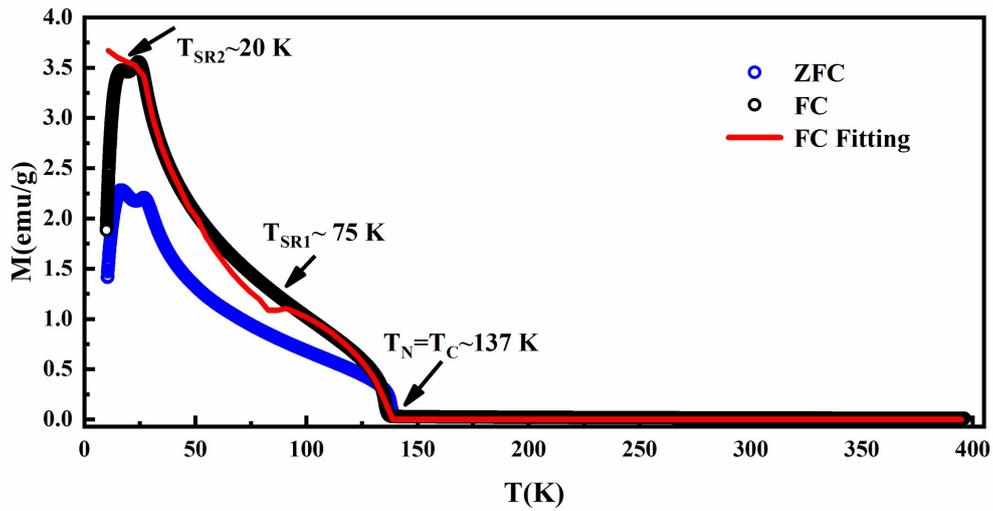

**Fig. 7.** Thermomagnetic curves of ErFe$_{0.1}$Cr$_{0.9}$O$_3$ at an external field of 100 Oe, and fitted with a molecular field model.

Figures 5 and 6 show the relationship between the magnetization and temperature of ErFe$_{0.9}$Cr$_{0.1}$O$_3$ and ErFe$_{0.1}$Cr$_{0.9}$O$_3$, respectively. Like most rare-earth perovskites, ErFe$_{1-x}$Cr$_x$O$_3$ exhibits multiple magnetic turning points at temperatures between 10 and 400 K. According to previous studies, we define them as $T_C$ (Curie temperature point), $T_N$ (Néel-like temperature point), $T_{SR1}$ (first spin reorientation point), $T_{SR2}$ (second spin reorientation temperature point).

In ErFe$_{0.9}$Cr$_{0.1}$O$_3$, no obvious Curie temperature point is observed, and even after 300 K, the system still undergoes an antiferromagnetic to paramagnetic transition phase at a slow rate. At about 105 K, due to the Dzialoshinski–Moriya (DM) interaction, the Fe$^{3+}$/Cr$^{3+}$ spins slightly deviate from the antiparallel angle, which finally leads to the appearance of weak ferromagnetism. The first spin reorientation occurs as the temperature cools further. That is, the spin planes of Fe$^{3+}$ and Cr$^{3+}$ gradually transition from $\Gamma_4(G_x, A_y, F_z; F_z^R)$ to $\Gamma_1(A_x, G_y, C_z; C_z^R)$ and end at the second reorientation temperature point. This conclusion can be well verified in many previous works [24,25].

Like ErFe$_{0.9}$Cr$_{0.1}$O$_3$, ErFe$_{0.1}$Cr$_{0.9}$O$_3$ undergoes a reorientation process between 75 K and 20 K, but the difference is that ErFe$_{0.1}$Cr$_{0.9}$O$_3$ does not undergo an antiferromagnetic to weak ferromagnetic transition until 137 K (for the same reason as previously analyzed), and at A weak ferromagnetic to paramagnetic phase transition occurs simultaneously. The main reason for the different $T_N$ may be that the magnetization temperature dependence of Cr$^{3+}$ is different from that of Fe$^{3+}$, and it is more sensitive.

Below the second reorientation temperature point ($T_{SR2}$), there is a large reversal of the magnetization. This is more than simple redirection can explain. Based on previous research on other rare-earth perovskites, DM interactions were found to surge as the temperature dropped. This may be the reason for the rapid magnetization reversal at low temperature [26].

To better describe the magnetization nature of the ErFe$_{1-x}$Cr$_x$O$_3$ system, we describe it by a simple four-lattice molecular field theory. That is, the sublattice of Fe$^{3+}$ and Cr$^{3+}$ in a and b sites is defined as L$_{Fe}^a$, L$_{Fe}^b$, L$_{Cr}^a$, L$_{Cr}^b$, the sublattice of rare earth ions in A site is defined as L$_{Er}$, and the sublattice of the remaining Cr$^{3+}$/Fe$^{3+}$ is defined as L$_{Cr}$/L$_{Fe}$. Because the magnetization data at ultra-low temperature is not obtained, and the Er$^{3+}$-Er$^{3+}$ interaction at high temperature is very small and negligible, we did not consider the Er$^{3+}$-Er$^{3+}$ superexchange. The mean field of each sublattice of ErFe$_{0.9}$Cr$_{0.1}$O$_3$ can be expressed as:

$$H_{Cr} = \lambda_{CrFe^a}M_{Fe^a} + \lambda_{CrEr}M_{Er} + h \#(3)$$

$$H_{Fe^b} = \lambda_{Fe^bFe^a}M_{Fe^a} + \lambda_{Fe^bEr}M_{Er} + h \#(4)$$

$$H_{Fe^a} = \lambda_{Fe^aFe^b}M_{Fe^b} + \lambda_{Fe^aCr}M_{Cr} + \lambda_{Fe^aEr}M_{Er} + h \#(5)$$

$$H_{Er} = \lambda_{ErCr}M_{Cr} + \lambda_{ErFe^a}M_{Fe^a} + \lambda_{ErFe^b}M_{Fe^b} + h \#(6)$$

For ErFe$_{0.1}$Cr$_{0.9}$O$_3$, the mean-field can be expressed as:

$$H_{Fe} = \lambda_{FeCr^a}M_{Cr^a} + \lambda_{FeEr}M_{Er} + h \#(7)$$

$$H_{Cr^b} = \lambda_{Cr^bCr^a}M_{Cr^a} + \lambda_{CrEr}M_{Er} + h \#(8)$$

$$H_{Cr^a} = \lambda_{Cr^aCr^b}M_{Cr^b} + \lambda_{Cr^aFe}M_{Fe} + \lambda_{Cr^aEr}M_{Er} + h \#(9)$$

$$H_{Er} = \lambda_{ErFe}M_{Fe} + \lambda_{ErCr^a}M_{Cr^a} + \lambda_{ErCr^b}M_{Cr^b} + h \#(10)$$

Where $\lambda_{ij}$ represents the molecular field constant between $i$ and $j$ sublattices, and it is proportional to the exchange constant $J_{ij}$, $M_i$ is the magnetization of $i$ sublattice, h is the external field. The magnetization of $i$ sublattice is:

$$M_i = X_i N_A g \mu_B S_i B_{si} \left( g \mu_B S_i H_i / k_B T \right) \#(11)$$

Where $x_i$ is the molar quantity of $i$ ions, $N_A$ is the Avogadro constant, $g$ is the lande factor, and $\mu_B$ represents the Bohr magneton, $S_i$ is the spin quantum number of $i$ ions ($S_{Fe}$ = 5/2, $S_{Cr}$ = 3/2, $S_{Er}$ = 3/2). The exchange constant $J_{MM}$ between $M$ and $M$ ions can be calculated by:

$$|J_{MM}| = 2 Z_{MM} S_M (S_M + 1) / 3 k_B T_N^M \#(12)$$

Where $Z_{MM}$ is the number of M ions required to be M ions nearest neighbours. $k_B$ is the Boltzmann constant, and $T_N^M$ is the phase transition temperature of ErMO$_3$. The exchange constants between Fe and Fe ($J_{Fe\text{-}Fe}$) and Cr and Cr ($J_{Cr\text{-}Cr}$) have been calculated to be 5.2 K and 25.0 K ($T_C$ of ErCrO$_3$~130 K, $T_C$ of ErFeO$_3$ ~ 625 K) [11,27]. Using formula (3) to (11) concurrently, the magnetization at each temperature can be calculated formula (11). To fit the experimental data using molecular field theory, we employ the most advanced heuristic algorithm—the Marine Predator Algorithm (MPA) [28].

However, the data obtained simply by the molecular field model are very different from the experimental values, because the molecular field model is only a simplified model. In order to obtain more accurate theoretical data, we not only consider the reorientation of Fe$^{3+}$/Cr$^{3+}$, but also the spin change of Er$^{3+}$ at the A site. To illustrate the effect of spin reorientation on the magnetism of ErFe$_{1-x}$Cr$_x$O$_3$, we treat the magnetization of ErFe$_{1-x}$Cr$_x$O$_3$ as a vector superposition of Fe$^{3+}$/Cr$^{3+}$ and Er$^{3+}$ ions. Since the A and B sites have different easy axis orientations, their interactions can be determined by their spin projections on specific planes.

Here, we define the effective spin $S_{eff}$ of $Er^{3+}$ as the projection of the full $Er^{3+}$ spin on the $Fe^{3+}/Cr^{3+}$ spin plane. The effective magnetic moment $S_{eff}$ (0.0001-1.5) is fitted by the best exchange constants $J_{Fe-Cr}$, $J_{Er-Fe}$, $J_{Er-Cr}$ obtained by fitting the high temperature section before (refer to Table 3). The final simulated data is represented by the red line in Figures 6 and 7. It can be seen that the results can basically cover the original experimental data well, and it is difficult to show a rapid magnetization transition only in the low temperature part. This is because we did not consider the super-strong DM interaction at low temperature. The effective spin and angle of $Er^{3+}$ corresponding to each temperature point can be seen in Figure 8. In the figure, we can observe that the effective spin change between the two reorientation temperature points (~25 K and ~75 K) is continuous and almost linear with temperature cooling.

**Table. 3.** Exchange constants of $ErFe_{0.9}Cr_{0.1}O_3$ and $ErFe_{0.1}Cr_{0.9}O_3$

| Sample | $J_{Fe-Fe}$ (K) | $J_{Cr-Cr}$ (K) | $J_{Fe-Cr}$ (K) | $J_{Er-Fe}$ (K) | $J_{Er-Cr}$ (K) |
|---|---|---|---|---|---|
| $ErFe_{0.9}Cr_{0.1}O_3$ | -25.000 | -4.643 | -8.317 | -1.279 | -8.994 |
| $ErFe_{0.1}Cr_{0.9}O_3$ | -25.000 | -4.643 | -8.091 | -1.256 | -7.959 |

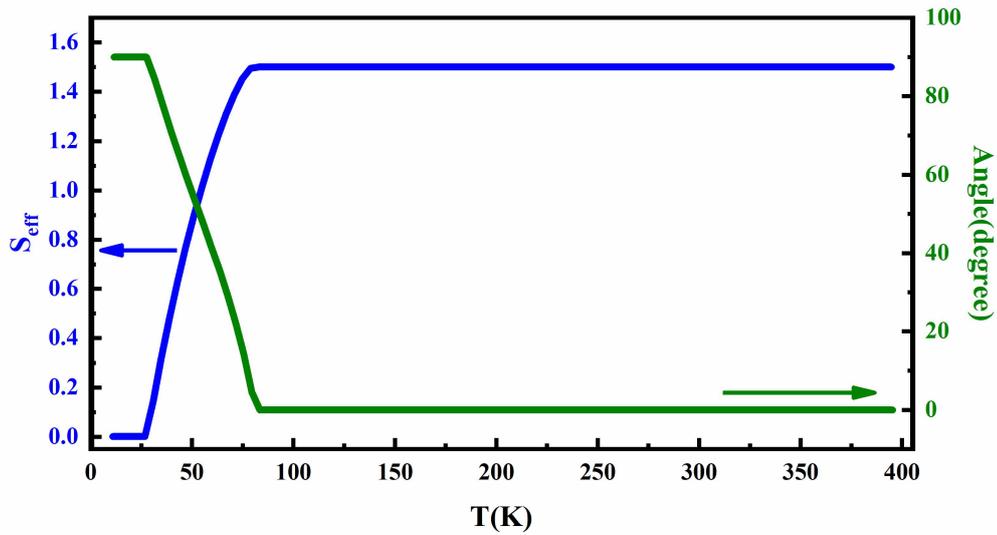

**Fig. 8.** The effective spin and angle of $Er^{3+}$ corresponding to each temperature point.

The results can well show that when reorienting, not only $Fe^{3+}/Cr^{3+}$ rotates on the a, c easy axis plane, but also the $Er^{3+}$ spin also rotates in the plane perpendicular to the a, c easy axis plane, the maximum The included angle is approximately 90° corresponding to Seff = 0. This can be attributed to the fact that these tilting moments of $Fe^{3+}/Cr^{3+}$ induce an overall internal magnetic field at the $Er^{3+}$ site, which in turn changes the orientation of $Er^{3+}$ spins in the direction of this induced field, i.e. the resulting effective magnetic moment [17].

## 4. Conclusions

In conclusion, the polycrystalline rare-earth perovskite $ErFe_{1-x}Cr_xO_3$ prepared by the sol-gel method in this report possess very complex magnetic characteristics. We determined the macroscopic magnetism at room temperature by magnetization curves and Mössbauer spectra. Accurate exchange constants are obtained by fitting the thermomagnetic curves of the system with molecular field theory. Interestingly, we found that in addition to the commonly observed $Fe^{3+}/Cr^{3+}$ spin reorientation, we found that the $Er^{3+}$ spin also changed with temperature during the general reorientation process. And fitting the regular $Er^{3+}$ spin on the $Fe^{3+}/Cr^{3+}$ spin plane projection, namely the effective spin. Reasonable explanation for the existence of errors. This study will guide researchers of rare earth perovskites to further understand the spin dynamics to a certain extent, and lay the foundation for future fast and precise spin-controlled devices.

## 5. Acknowledgments

This work was partially supported by the National Natural Science Foundation of China (grant number 12105137), the National Undergraduate Innovation and Entrepreneurship Training Program Support Projects of China, the Natural Science Foundation of Hunan Province, China (grant number S202110555177), the Natural Science Foundation of Hunan Province, China (grant number 2020JJ4517), Research Foundation of Education Bureau of Hunan Province, China (grant number 19A433, 19C1621).

## 6. Conflicts of interest

There are no conflicts of interest to declare.

2020, **152**, 113377.